\documentclass[aps,prb,10pt,superscriptaddress,longbibliography]{revtex4-2}

\usepackage{tikz}
\usepackage[cmex10]{amsmath}
\usepackage{amssymb}
\usepackage{physics}
\usepackage{upgreek}
\usepackage{titlesec}
\usepackage{siunitx}
\usepackage{here}
\usepackage{tabularx}
\usepackage{color}
\usepackage{comment}
\usepackage{xcolor}
\usepackage{ulem}
\newcommand{\add}[1]{\textcolor{black}{#1}}
\newcommand{\erase}[1]{}
\newcommand{\addt}[1]{\textcolor{black}{#1}}
\newcommand{\eraset}[1]{}

\usepackage{hyperref}
\hypersetup{
    colorlinks=true,
    linktocpage=true,
    linkcolor=blue,
    filecolor=magenta,      
    urlcolor=cyan,
    bookmarks=true,}

\graphicspath{{images/}}

\begin{document}
\title{Efficient spectrum analysis for multi-junction superconducting circuit}

\author{A. Tomonaga}\email{akiyoshi.tomonaga@aist.go.jp}
\affiliation{Institute of Advanced Industrial Science and Technology (AIST), Tsukuba, Ibaraki 305-8563, Japan}
\affiliation{Department of Physics, Tokyo University of Science, 1--3 Kagurazaka, Shinjuku, Tokyo 162--0825, Japan}
\affiliation{RIKEN Center for Quantum Computing (RQC), 2--1 Hirosawa, Wako, Saitama 351--0198, Japan}
\author{H. Mukai}
\affiliation{Department of Physics, Tokyo University of Science, 1--3 Kagurazaka, Shinjuku, Tokyo 162--0825, Japan}
\affiliation{RIKEN Center for Quantum Computing (RQC), 2--1 Hirosawa, Wako, Saitama 351--0198, Japan}
\author{K. Mizuno}
\affiliation{Institute of Advanced Industrial Science and Technology (AIST), Tsukuba, Ibaraki 305-8563, Japan}
\author{J. S. Tsai}
\affiliation{Department of Physics, Tokyo University of Science, 1--3 Kagurazaka, Shinjuku, Tokyo 162--0825, Japan}
\affiliation{RIKEN Center for Quantum Computing (RQC), 2--1 Hirosawa, Wako, Saitama 351--0198, Japan}

\begin{abstract}
The extraction of transition frequencies from a spectrum has conventionally relied on empirical methods, and particularly in complex systems, it is a time-consuming and cumbersome process.
To address this challenge, we establish a semi-automated efficient and precise spectrum analysis method.
It first employs image processing methods to extract transition frequencies, subsequently estimating the Hamiltonians of a superconducting quantum circuit containing multiple Josephson junctions.
In addition, we determine the suitable range of approximations in simulation methods, evaluating the physical reliability of the analysis.

\end{abstract}

\maketitle

\section{Introduction}
\begin{figure*}[b!]
\centering
\includegraphics[width=180mm]{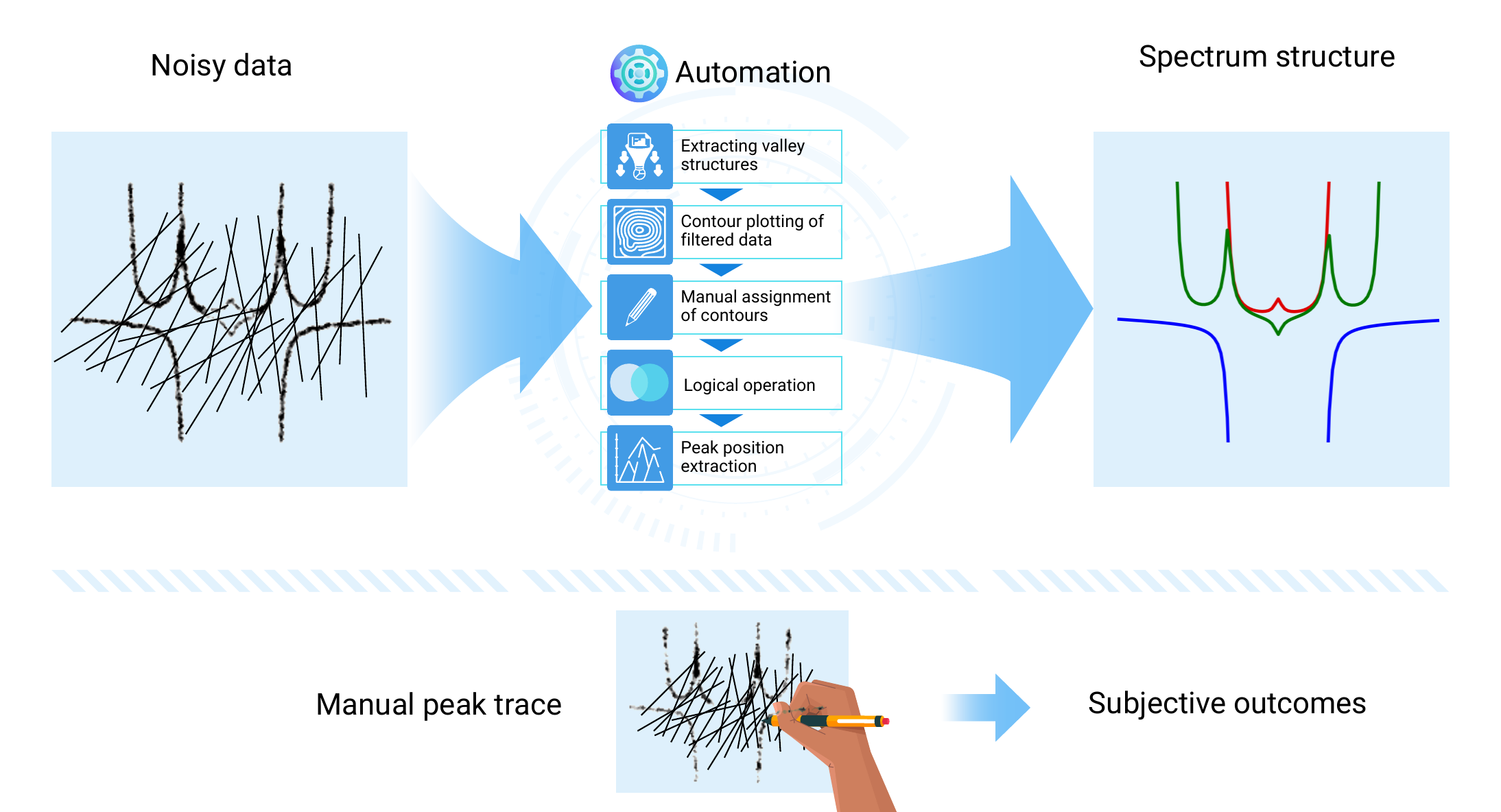}
\caption{Image of peak trace and flow of the peak trace program.}
\label{FC}
\end{figure*}
Superconducting quantum circuits have garnered significant interest as one of the most important candidates for quantum information processing and as a versatile platform for a wide range of fundamental studies on microwave quantum optics~\cite{nakamura_coherent_1999,gu_microwave_2017,blais_circuit_2021,kwon_gate-based_2021}. Recently, several unique quantum phenomena, \erase{(e.g.,}\add{such as} waveguide quantum electrodynamics, ultrastrong coupling, and fluxonium exploration\erase{)}, have been observed~\cite{sheremet_waveguide_2023,forn-diaz_ultrastrong_2019,PhysRevX.9.041041}.
In these \add{complex}\erase{new} experiments and research on integrated circuits, which are increasingly handling large amounts of data, there is a growing demand for efficiency and automation in data analysis.

In research on superconducting quantum circuits, measurements often begin by observing the energy spectrum as a function of the magnetic fields or microwave power and comparing the results with theoretical models. 
Analyzing the spectrum requires extracting a series of data points corresponding to the individual transition energies of the system from the experimental data and applying a fit.
Compared to circuits with weak nonlinearity, such as transmons~\cite{schreier_suppressing_2008,lu_characterizing_2021}, the spectra of ultrastrongly coupled systems or fluxonium circuits exhibit highly nonlinear features~\cite{earnest_realization_2018,mencia_integer_2024,tomonaga_quasiparticle_2021,yoshihara_superconducting_2017,niemczyk_circuit_2010}.
Extracting peak points from these spectra is very challenging and time-consuming because the spectra vary in terms of the number of transition frequencies, crowding of different frequencies, and height of the peaks (depending on the location).
Nevertheless, the spectral structure of superconducting circuits can often be predicted precisely using the circuit Hamiltonian.
Using these predictions, the transition frequencies can be extracted from the measured spectrum. However, owing to the presence of noise, this process is not straightforward. 

When the signals exhibit simple structures and are sufficiently stronger than the noise, the maxima (or minima) can be identified for each row (or column), and the resonance points can be determined via fits to relations such as the Lorentzian function. 
However, when multiple transition frequencies are in close proximity to each other, dividing the spectrum into smaller regions is necessary.
Additionally, when the signal-to-noise ratio (SNR) is low, conventional methods, such as Lorentzian fitting and minimum or maximum point extraction, often fail.

\begin{figure}[t]
\centering
\includegraphics[width=180mm]{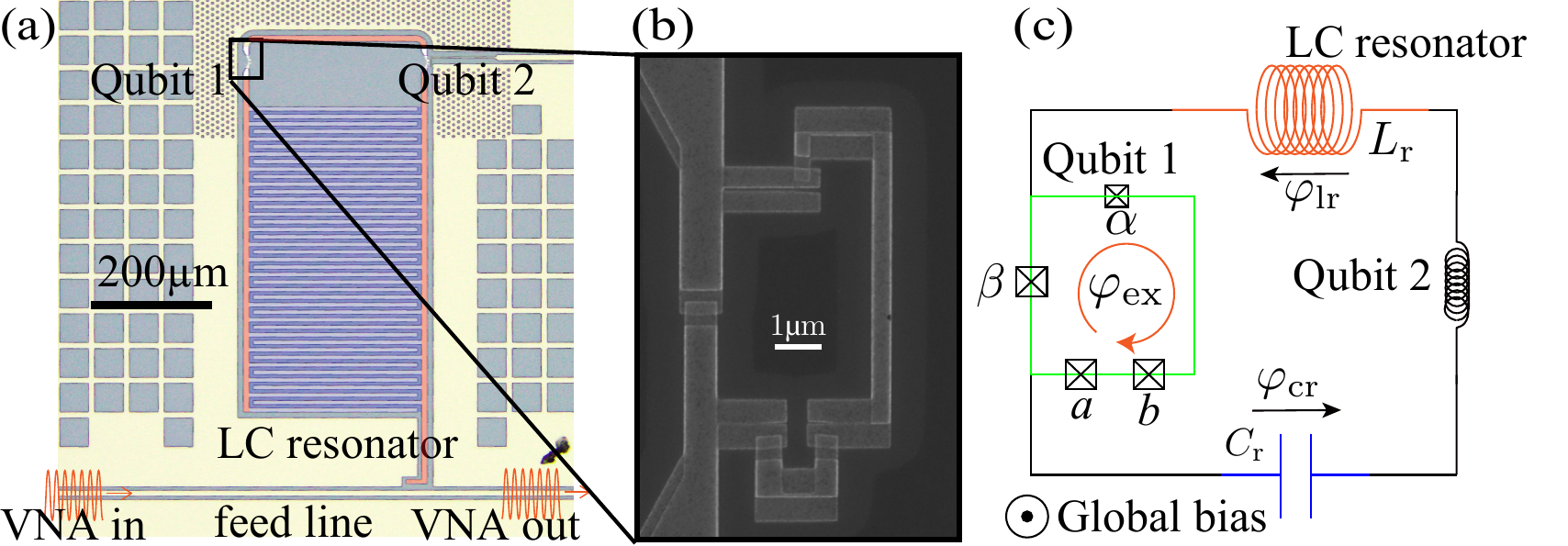}
    \caption{
    (a) \add{False-color} optical microscopy image of measured superconducting circuit. 
    The LC resonator \erase{was}\add{is} fabricated using 50-nm Nb thin film.
    The spectrum \erase{was}\add{is} obtained by measuring $S_{21}$ through the feed line using a vector network analyzer (VNA). 
    The qubit energy \erase{was}\add{is} tuned by a magnetic coil that generated a uniform magnetic field around the whole chip. 
    (b) Scanning electron microscope (SEM) image of the flux qubit coupled to the LC resonator via a $\beta$-junction. 
    (c) Schematic \erase{image }of the measured superconducting circuit (the names of the circuit components are defined). \add{The symbols $\alpha, \beta, a,$ and $b$ represent the labels of the junctions and their relative size ratios, where $a$ is taken as 1.} \add{The colors of the inductor and capacitor in the resonator match those in panel (a).}
    Qubit 2 is treated as a classical inductor~\cite{harris_compound_2009}. The circuit design follows the design rules introduced in Ref.~\onlinecite{tomonaga_quasiparticle_2021}, which ensure low sensitivity to charge fluctuations for \add{the} four junction\erase{s in a} flux qubit.
    }
\label{Circuit}
\end{figure}
\begin{figure*}[t!]
\centering
\includegraphics[width=180mm]{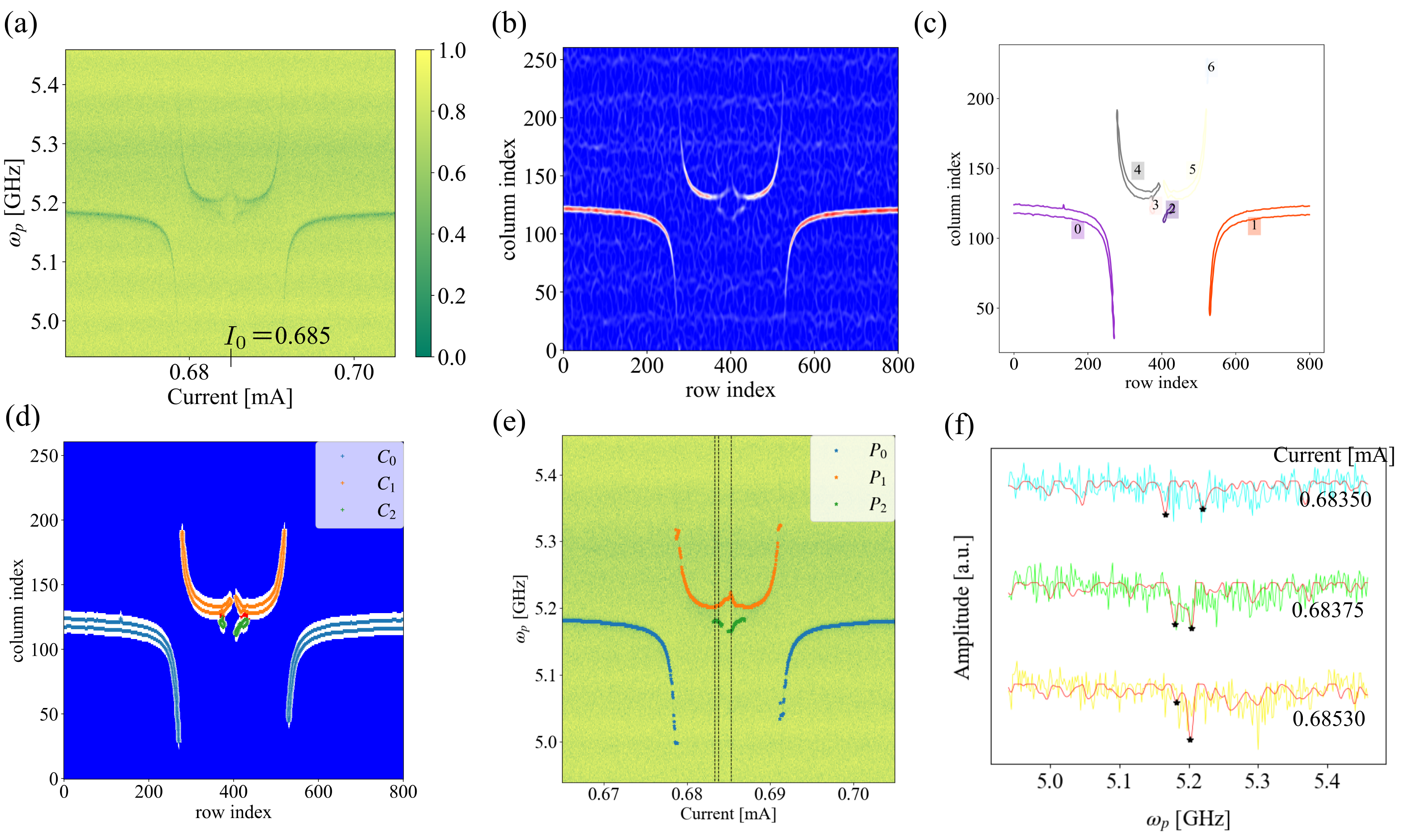}
    \caption{
    (a) Example of a superconducting circuit spectrum. The input power for the chip is approximately -130 dBm, and the intermediate frequency (IF) is 100 Hz according to a vector network analyzer (VNA).
    (b) Filtered image of the spectrum, constructed via multi-scale filter methods to emphasize the \erase{ride}\add{valley} structure.
    (c) Extracted contour lines using a threshold with an amplitude of 0.25 and a line length of 20 points.
    (d) Image in which contour lines are given a width and an exclusive logical (XOR) operation is applied for each \add{pair of traced regions}\erase{transition frequency}.
    (e) Traced peaks of the spectrum. 
    (f) Signal sliced for each \add{dashed lines} shown in panel (e). 
    Red curves represent filtered signal in panel (b). Not all slices show peak points correspond\add{ing}\erase{s} to the \erase{maximum}\add{minimum}. 
    }
\label{PT}
\end{figure*}

In such cases, one remaining approach is for researchers to manually trace lines on a two-dimensional color map by hand. However, this method is highly subjective and time-consuming. Even if a subjective approach yields parameters that reproduce experimental results, automation is essential for handling \eraset{large datasets}\addt{dozens spectra at once} and conducting complex analyses.

Therefore, in this paper, we develop a method for extracting characteristic structures from highly nonlinear spectra using image processing techniques, such as visual identification of \erase{ridges}\add{valleys or ridges}, to efficiently obtain the data points corresponding to each transition energy.
We also provide a\add{n} effective process \erase{of}\add{for} fitting a spectrum to a circuit model with multiple Josephson junctions.
As a specific example, we demonstrate an efficient fitting technique that obtains the circuit parameters from the spectrum of an ultrastrongly coupled superconducting quantum circuit.

Moreover, when representing a physical system using a finite approximation, the expansion terms must be taken until the solution adequately converges. 
In the latter part of this paper, we verify the convergence of eigenvalues and provide a\add{n} understanding suitable ranges of the charge space of qubit and the Fock space of the resonator.
In the circuit model calculations, the coupling term with the resonator is expanded using eigenvectors of the qubit. 
This approach reduces computational time and enables efficient analysis\add{;}\erase{, and} we also discuss the convergence of solutions in this context.
Finally, we examine the validity of the parameters obtained from fitting the circuit model.

The remainder of this paper is organized as follows. Section II describes the process of peak tracing, Section III details the Rabi model fit, Section IV explains the circuit model fit, Section V discusses the results, and Section VI summarizes the conclusions of the study.

\section{Peak Tracing}\label{sec:Peak}
The measured circuit is composed of flux qubits~\cite{orlando_superconducting_1999} and a resonator, as shown in Fig.~\ref{Circuit}(a). 
They are ultrastrongly coupled.
The spectrum [Fig.~\ref{PT}(a)] is experimentally obtained by varying the global magnetic field in the region where the qubit frequency is close to the resonator frequency\erase{ since signal is not filter out}. 
The transmission signal $S_{21}$ is measured using the in and out ports of a vector network analyzer (VNA), as shown in Fig.~\ref{Circuit}(a).
To explain a process of obtaining the peak trace, the measured spectrum shown in Fig.~\ref{PT}(a) is used as an example for analysis.
The processed spectra obtained after each step are shown in Figs.~\ref{PT}(b)-(e).
The VNA's intermediate-frequency (IF) bandwidth is set to 100 Hz to reduce the measurement time, resulting in a relatively low SNR for this spectrum [Fig.~\ref{PT}(f)].

The spectra of devices such as superconducting quantum circuits are obtained by measuring the intensities and phases of the transmitted or reflected signals.
\erase{An} \add{Signal intensity of an} energy transition \erase{signal }exhibits a Lorentzian shape.
Consequently, spectrum often show the form of a \erase{ridge or crest}\add{valley or ridge} structure.
The spectrum used in this study shown in Fig.\ref{PT}(a) depicted the \erase{ridge}\add{valley} structure.

\begin{figure*}[t]
\centering
\includegraphics[width=180mm]{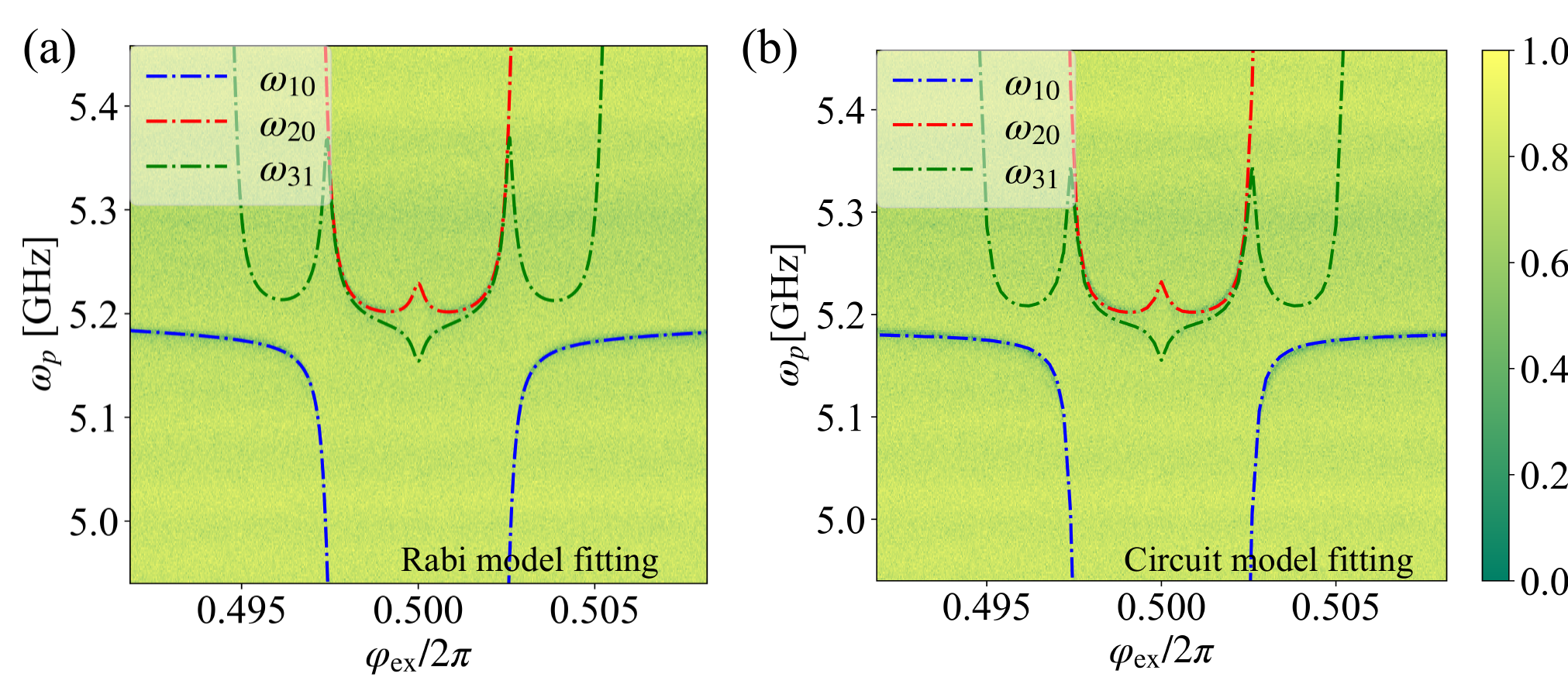}
    \caption{
    (a) Measured spectrum fitted by the Rabi model with parameters $g/2\pi = 3.45~\mathrm{GHz}$, $\Delta/2\pi = 0.83$ GHz, $\omega_\mathrm{r}/2\pi = 5.17~\mathrm{GHz}$, $A_- = 0.13\erase{4.01\times10^{-4}}$, $A_+=0.11\erase{3.42\times10^{-4}}$, and $I_p$=323 nA.
    (b) Measured spectrum fitted by the circuit model with parameters $E_\mathrm{J}/2\pi = 278~\mathrm{GHz}$, $E_c/2\pi = 2.88~\mathrm{GHz}$, $\alpha=0.66$, $\beta = 1.62$, $L_\mathrm{r} = 5.00$ nH, and $C_\mathrm{r} = 175$ fF.}
\label{Fit}
\end{figure*}

The peak tracing procedure \erase{was}\add{is} as follows:
\begin{enumerate}
    \item \textit{Extracting \erase{ridge}\add{valley} structures from noisy spectra.} \\
    To remove background noise, the multi-scale line filter processing method~\cite{sato_three-dimensional_1998,walt_scikit-image_2014}, which is demonstrated in Figs.~\ref{PT}(a) and (b), \erase{was}\add{is} used. 
    This filter \erase{was}\add{is} originally developed to identify blood vessels in magnetic resonance imaging (MRI)~\cite{sato_three-dimensional_1998}, the appearance of which resembles the spectrum of superconducting circuits.
    By appropriately selecting the scale factor to recover various line structures, even complex \erase{ridge}\add{valley} such as the $\omega_{31}$ and $\omega_{20}$ transitions, which are very close to each other \add{[}as shown in Fig.~\ref{PT}(b) and Fig.~\ref{Fit}\add{]}, can be separated and extracted.

    \item \textit{Contour plotting of filtered data.} \\
    Figure~\ref{PT}(c) displays contours of the filtered data shown in Fig.~\ref{PT}(b). 
    The iso-valued contours \erase{were fund}\add{are found} by performing the marching squares method in a two dimensional array~\cite{lorensen_marching_1987, walt_scikit-image_2014}.
    The contour length and height are specified as the thresholds by which the \erase{ridge}\add{valley} structures are extracted. 
    While transition spectra typically form elongated ellipses or arc shapes, noise usually appears as small circular shapes.
    Therefore, by setting a threshold for the contour length, noise would be removed well.
    
    \item \textit{Assignment of contours to transition frequencies.}\\
    The extracted contours labeled in Fig.~\ref{PT}(c) are manually assigned to transition frequencies one by one. 
    For example, in Figs.~\ref{PT}(c) and (d), each contour line is assigned to three contours group ($C_k$): $C_0 = \lbrace \erase{1}\add{0}, \erase{2}\add{1} \rbrace$, $C_1 = \lbrace 4, 5 \rbrace$, and $C_2 = \lbrace2, 3 \rbrace$, where the numbers in braket correspond to the labels of the contour lines and others are ignored as noise.

    \item \textit{Logical operation for overlapping area.}\\
    Several points in the vertical and horizontal directions are assigned to each extracted contour as areas that include the transition frequencies shown in Fig.~\ref{PT}(d). 
    In cases where \erase{ridges}\add{valleys} are close to each other, neighboring \erase{ridges}\add{valleys} may be erroneously interpreted as a peak point. 
    \add{To ensure that these regions do not overlap, an exclusive logical (XOR) operation is applied to all pairs of extracted regions. The red-colored areas in Fig.~\ref{PT}(d) indicate the segments eliminated by the XOR operation.}
    \erase{Figure~\ref{PT}(d) depicts the application of an exclusive logical (XOR) operation, which ensures that these regions do not overlap.}

    \item \textit{Peak position extraction.}\\
    The transition frequencies that give the minimum value in the regions of the filtered spectrum shown in Fig.~\ref{PT}(b) is extracted. 
    Thus, we obtain groups of peak points (each transition frequencies) $P_k$ shown in Fig.~\ref{PT}(e).
    If the noise level is not excessively high, Lorentzian fitting along the horizontal axis can also be useful for determining the peak values.
    
\end{enumerate}

Figure~\ref{PT}(f) shows the $S_{21}$ signal and extracted peak points (marked by stars) for several bias points indicated \erase{at the top of}\add{by dotted lines on} Fig.~\ref{PT}(e). 
As shown in the figure, the peak values \erase{did }not always coincide with the global minimum. 
Although it is difficult identifying the spectrum's peak points from $S_{21}$ at each bias point, the three-dimensional plot and image processing method enabled efficient and straightforward peak extraction.
In the proposed method, subjectivity is involved only in the step to classify the contours into each transition frequency. 
Meanwhile, all other steps are processed automatically.

\section{Rabi model fit}
When the computational cost of the model you would like to use in the final is expensive, starting with a simpler fitting model would reduce the cost.
When a superconducting resonator is coupled to a flux qubit (in a fashion similar to that of our circuit), the two-level approximation is useful.
The coupled system of a two-level qubit and resonator, including the ultrastrong~\cite{niemczyk_circuit_2010,tomonaga_spectral_2025} and deep-strong~\cite{yoshihara_superconducting_2017,forn-diaz_ultrastrong_2017} couplings, is properly described by the Rabi model Hamiltonian $\mathcal{H}_\mathrm{Rabi}$, which represents the phenomenological coupling between the spins and electromagnetic field. The Rabi model Hamiltonian is expressed as
\begin{align}
 \mathcal{H}_\mathrm{Rabi} = 
    \frac{\hbar}{2}\qty(\varepsilon\hat{\sigma}_\mathrm{z}+\Delta\hat{\sigma}_\mathrm{x}) + \hbar\omega_\mathrm{r} \hat{a}^\dagger \hat{a} + \hbar g\hat{\sigma}_\mathrm{z}\qty(\hat{a}^\dagger + \hat{a})\,,
    \label{Rabi}   
\end{align}
where $\hat{\sigma}$ represents the Pauli operator and $g$ is the coupling constant between the qubit and the resonator.
In the flux qubit, $\varepsilon$ represents the external field term applied to the spin corresponding to the energy of the loop current. In addition, $\Delta$ denotes the energy gap of the qubit and is exponentially proportional to $\alpha$ \add{defined in Fig.~\ref{Circuit}(c)}~\cite{schwarz_gradiometric_2013}.

In the plot illustrated in Fig.~\ref{PT}(a), the data are expressed as a function of current because the spectrum \erase{was}\add{is} \erase{derived}\add{obtained} by introducing flux into the qubit loop using a room-temperature bias current source. Considering the current bias, the \erase{actual }fitting function is given by \cite{tomonaga_spectral_2025}
\begin{align}
 \mathcal{H}'_\mathrm{Rabi} =& 
    \frac{\hbar}{2}\qty{\tilde{\varepsilon}(I-I_0)\hat{\sigma}_\mathrm{z}+\erase{\hbar}\Delta\hat{\sigma}_\mathrm{x}} + \hbar\omega_\mathrm{r}\{1\pm A_{\pm}(I-I_0)\}\hat{a}^\dagger \hat{a} + \hbar g\hat{\sigma}_\mathrm{z}\qty(\hat{a}^\dagger + \hat{a})\,,
    \label{Rabi_exp}   
\end{align}
where $\hbar\varepsilon=\hbar\tilde{\varepsilon}(I-I_0)=2I_p\Phi_0(\varphi_\mathrm{ex}/2\pi-0.5)$, $I$ is the bias current from a room-temperature current source, $I_0$ is the current for $\varphi_\mathrm{ex}/2\pi=0.5$ \add{as shown in Fig~\ref{PT}(a)}, $\tilde{\varepsilon}$ is a fitting parameter, and $I_p$ is the qubit persistent current. 
In addition, the resonator frequency varies according to the flux because the resonator is coupled directly to the $\beta$\add{-}junction, and the coefficients $A_\pm$ are used as fitting parameters to account for this variation. These coefficients differ in value in the positive and negative directions relative to $\varepsilon = 0$, which is attributed to the slope caused by the presence of a neighboring qubit.

For the fit to the Rabi model Hamiltonian, we associate\erase{d} the data points $P_0$, $P_1$, and $P_2$ \add{[}shown in Fig.\ref{PT}(e)\add{]} with transition frequencies $\omega_{10}$, $\omega_{20}$, $\omega_{31}$, \add{each calculated from the eigenvalues of Eq.~\eqref{Rabi_exp}.}\erase{ respectively.} 
In the Rabi model, it is desirable to consider Fock states with more than 10 states in the ultrastrong coupling regime. However, the computational cost is significantly cheaper than that of circuit models.
The fitting results are shown in Fig.\ref{Fit}(a), which \erase{demonstrates the reproducibility of}\add{reproduce} the experimental data \add{well}, and obtained parameters are $g/2\pi = 3.45$ GHz, $\Delta/2\pi = 0.83$ GHz, $\omega_\mathrm{r}/2\pi = 5.17~\mathrm{GHz}$, $A_- = 4.01\times10^{-4}$, $A_+ = 3.42\times10^{-4}$, and $I_p = 323~\mathrm{nA}$.

\begin{figure*}[t]
\centering
\includegraphics[width=180mm]{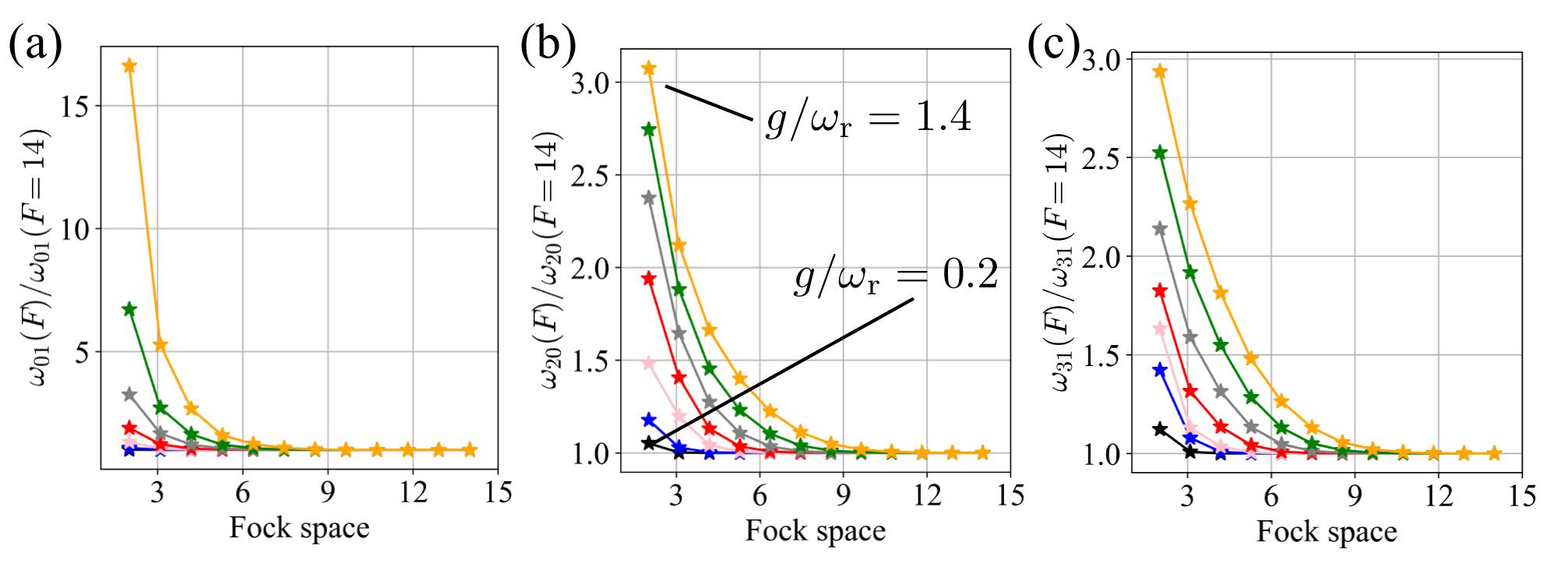}
    \caption{(a) Ratio of when each Fock space is taken along the horizontal axis to when the Fock space reaches up to 14 states (F represents the Fock space). The colors of the lines correspond to values of $g/\omega_\mathrm{r}$ ranging from 1.4 to 0.2, in steps of 0.2 and in descending order. The other parameters are $\Delta\add{/}\omega_\mathrm{r}=0.4$ and $\varepsilon=0$
    }
\label{Fock}
\end{figure*}

\begin{figure}[t]
\centering
\includegraphics[width=85mm]{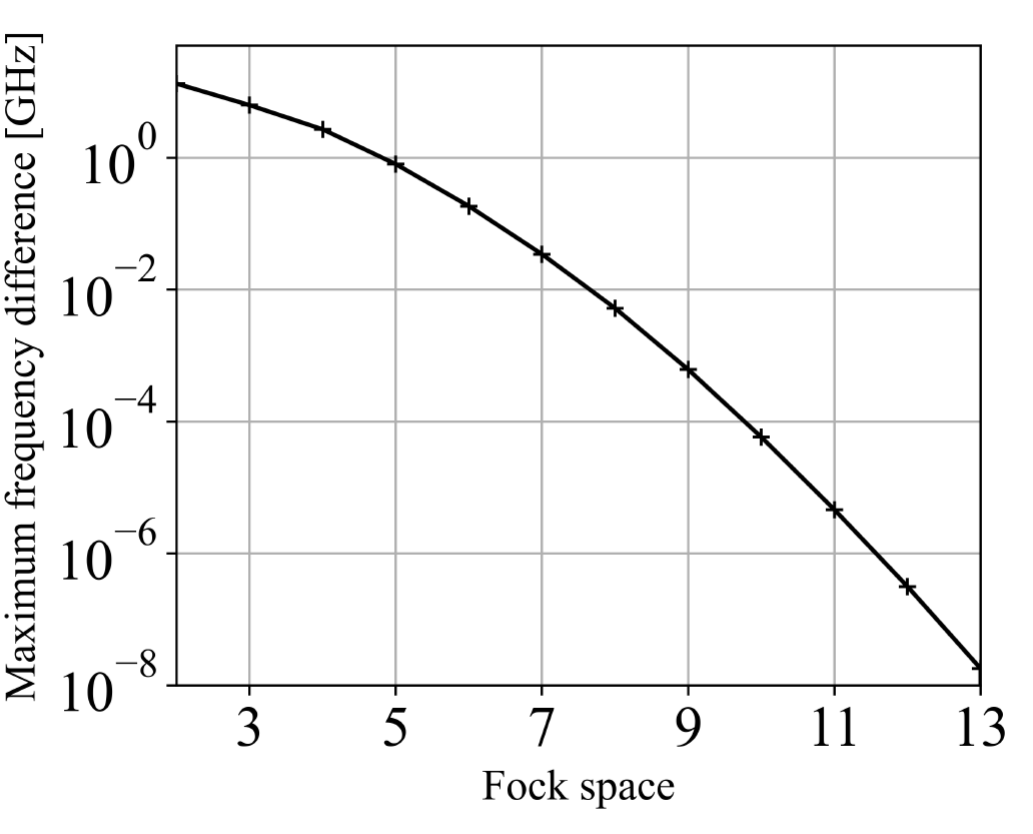}
\caption{\erase{(a)} Plot of \add{max$\{\abs{\omega_{10}(F)-\omega_{10}(14)}+\abs{\omega_{20}(F)-\omega_{20}(14)}+\abs{\omega_{31}(F)-\omega_{31}(14)}\}$} for the range $\varphi_\mathrm{ex}/2\pi=0.49-0.50$. Increasing the Fock space monotonically improves the accuracy. The graph shows Fock=9 is enough to the spectrum fitting.
}
\label{Fspace}
\end{figure}
We now discuss the validity of the approximations used in the Rabi models for the fitting.
During the numerical calculations, we used a finite number of photon states when expanding the system in terms of the wavefunctions of a harmonic oscillator.
For a spectrum in the limit of weak pumping light, it is sufficient to calculate only a few of the lowest excited states. 
However, as the coupling constant increases, a larger number of expansion terms become necessary.

Figure~\ref{Fock} shows the ratio of the transition frequencies calculated when the Fock space of the resonator ($F$) \erase{was}\add{is} extended up to 14 to those calculated when the Fock space \erase{was}\add{is} restricted to values between $F = 2$ and $F=13$, with $g/\omega_\mathrm{r}$ increased by increments of 0.2. 
The results indicate that, as the coupling constant increased, the values failed to converge unless the Fock space \erase{was}\add{is} sufficiently expanded. For $g/\omega_\mathrm{r} \approx 1$, including more than 9 Fock states is desirable.

Using the fitting parameters obtained, the maximum frequency differences between the eigenvalues at $F = 14$ and those at each Fock space for $\omega_{10}$, $\omega_{20}$, and $\omega_{31}$ is shown in Fig.~\ref{Fspace}. 
The Lorentzian-shaped dips forming the \erase{ridge}\add{valley} structure of the spectrum \erase{had}\add{have} a full width at half maximum (FWHM) of approximately 10 MHz. 
Thus, taking a Fock space \add{more than nine}\erase{of approximately 9}, where the maximum difference \erase{was}\add{is} below 1 MHz, \erase{was}\add{is} reasonable\erase{ for the noise in Fig.~\ref{Fspace}}. 
The fit \erase{was}\add{is} performed using $F =13$.

\section{Circuit model fit}
Here, we present the circuit Hamiltonian used for fitting. The detailed derivation is provided in~\ref{sec:A0}.
The total Hamiltonian of the circuit can be written as
\begin{align}
   \mathcal{H}_{\mathrm{QR}}
   =\mathcal{H}_\mathrm{Q}+\mathcal{H}_\mathrm{R}+\mathcal{H}_\mathrm{Int} \,.
    \label{eq:TotalHami_A}
\end{align}
The qubit Hamiltonian is found as
\begin{align}
\mathcal{H}_\mathrm{Q}=4E_\mathrm{c}{\hat{\vb{q}}}^\mathrm{T} \tilde{\vb{M}}^{-1}\hat{{\vb{q}}}
    +\mathcal{U}\,,
    \label{HQ13}
\end{align}
where $E_c$ is the charge energy of the $a$ -junction, $\hat{\bold{q}}=(\hat{q}_\beta,\hat{q}_\alpha, \hat{q}_a)^\mathrm{T}$ is the charge basis vector, $\tilde{\bold{M}}$ is the normalized mass matrix and $\mathcal{U}$ is the potential of the qubit. The basis $\hat{q}_b$ is eliminated using Kirchhoff's law of the qubit loop.
Using the creation and annihilation operators, the Hamiltonian of the resonator and the interaction between resonator and qubit are represented by
\begin{align}
  \mathcal{H}_\mathrm{r}=\hbar\omega_\mathrm{r}(\hat{a}^\dagger \hat{a}+\frac{1}{2})\,,
\end{align}
and
\begin{align}
  \mathcal{H}_\mathrm{int}=\frac{I_\mathrm{r}\Phi_0}{2\pi}\hat{\varphi}_\beta(\hat{a}^\dagger+\hat{a})\,,
\end{align}
respectively, where $\omega_\mathrm{r}=1/\sqrt{L_\mathrm{r}C_\mathrm{r}}$ is the bare frequency and $I_\mathrm{r}=\sqrt{\hbar/(2L_\mathrm{r}\sqrt{L_\mathrm{r}C_\mathrm{r}})}$ is the zero point fluctuation current of the resonator. The phase across \add{the} $\beta$-junction represents $\hat{\varphi}_\beta$.

\add{Because the Qubit Hamiltonian is represented by six order tensor, }when the resonator terms are calculated using the product state\erase{ of this sixth-order tensor}, the total tensor rank increases to eight, resulting in a high computational cost. 
Subsequently, after diagonalizing the qubit Hamiltonian, the coupling terms are expanded in the eigenstates of the qubit. 
This allows the calculation to be split into tensors of ranks six and four, thereby reducing the computational cost.
Using the eigenvector $\ket{i}_i$ (${i}\in \mathbb{N}$) of the qubit Hamiltonian $\mathcal{H}_\mathrm{Q}$, we obtain \add{the following}.
\begin{align}
\mathcal{H}_{\mathrm{tot}}=&\hbar\sum_{i}{\Omega_i \ket{i}\bra{i}}
+\mathcal{H}_\mathrm{r}+\hbar\sum_{i,j}{g_{ij}\ket{i}\bra{j}(\hat{a}^\dagger\!+\hat{a})}\,,
\label{Hket}
\end{align}
where $\hbar\Omega_{i}$ is the $i$-th eigenenergy of qubit and $\hbar g_{ij}=I_\mathrm{r}\Phi_0\mel{i}{\hat{\varphi}_{\beta}}{j}$ is the coupling matrix element.

Using this Hamiltonian, the circuit model is fitted to the curve of the Rabi model. Because the Rabi model precisely reproduces the experimental data and the noise is statistically removed during the Rabi fitting, it is possible to efficiently determine the circuit parameters by matching the circuit model to the Rabi model using fewer data points.
The results of the fitting using the circuit model are shown in Fig.~\ref{Fit}(b), which shows that the circuit model also reproduces the experimental data well.
And we obtain circuit parameters: $E_\mathrm{J}/2\pi=278 \mathrm{GHz}$, $E_c/2\pi=2.88 \mathrm{GHz}$, $\alpha=0.66$, $\beta=1.62$, $L_\mathrm{r}=5.00$ nH, and $C_\mathrm{r}=175$ fF.
\add{Here, we assume that junctions a and b have identical sizes.}

\begin{figure}[b]
\centering
\includegraphics[width=85mm]{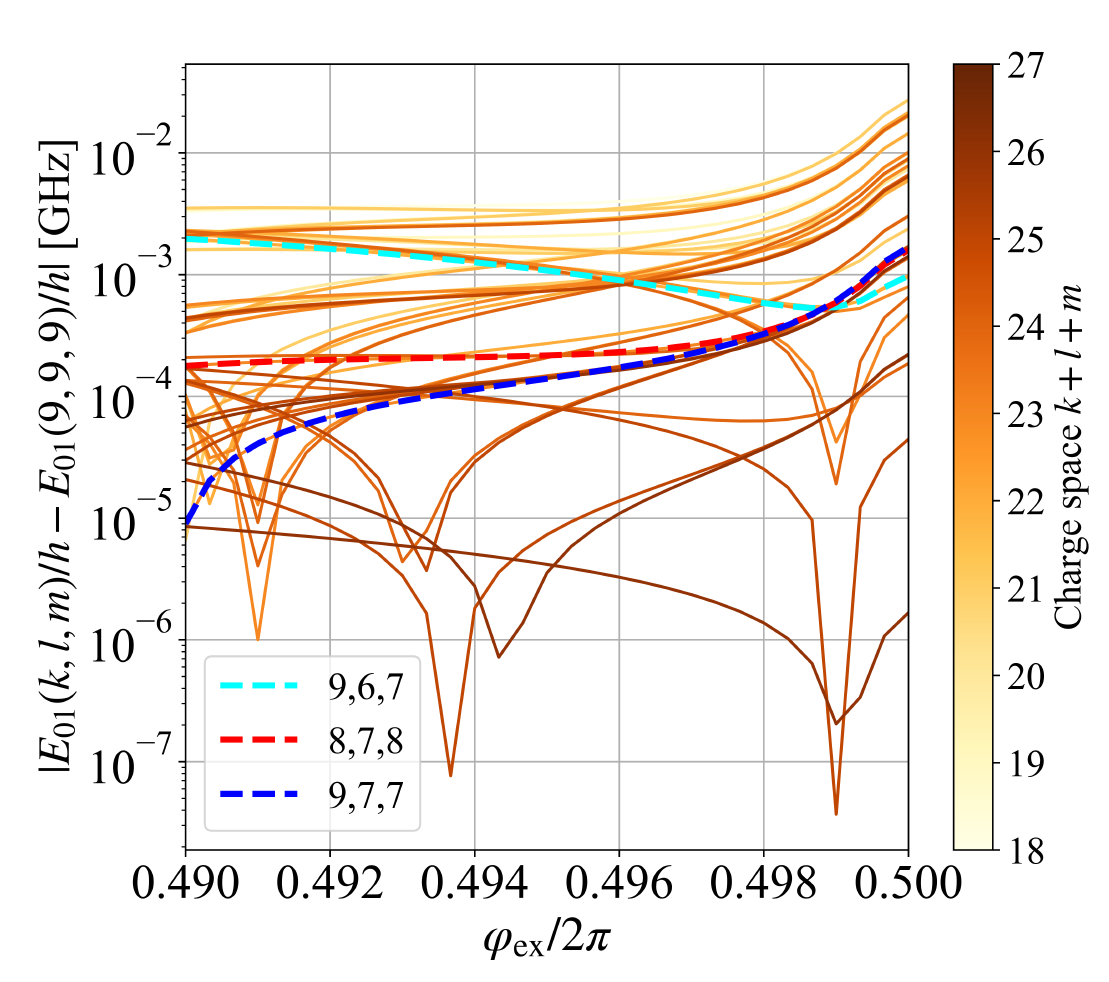}
    \caption{
    Transition frequency difference between the first excited state and the ground state of the qubit, calculated at each bias point as $\abs{E_{01}(k,l,m)/h - E_{01}(9,9,9)/h}$.
    As $k+l+m$ increases, that is, as the color becomes darker, the eigenvalues converge.
    The configurations $(k,l,m) = (9,6,7)$, $(8,7,8)$, and $(9,7,7)$ have relatively low computational costs and maintain errors below 2 MHz across all bias points.
    }
\label{charge}
\end{figure}
We now discuss the extent to which the charge number of the qubit Hamiltonian should be considered ~\cite{robertson_quantum_2006, yoshihara_hamiltonian_2022}.
As shown in Eq.~\eqref{wavefunc}, which follows from the Bardeen-Cooper-Schrieffer (BCS) theory, the superconducting state is expressed as a superposition of plane waves.
The corresponding wave numbers match the number of Cooper pairs \erase{and correspond to the values of $k$, $l$, and $m$ in Eq.~\eqref{H1tensor}}\add{in the charge basis $\hat{{\vb{q}}}$~(see~\ref{sec:A0})}.
Additionally, $\{k,~l,~m\} \in \mathbb{N}$ are related to the $\beta$, $\alpha$, and $a$ junctions, respectively. 

Considering the circuits shown in Fig.~\ref{Circuit}(c) and the spectrum shown in Fig.~\ref{PT}(a), we only need to consider the transition energy between the ground and first excited states $E_{01}$ of the qubit.
Figure~\ref{charge} shows the differences in $E_{01}$ at each bias point for $(k, l, m) = (6, 6, 6)$ and the subsequent values of $k, l, m$ compared to the case in which $(k, l, m) = (9, 9, 9)$. 
These results show\erase{s} that as $k$, $l$, and $m$ increased, the system converged, not behaving monotonically.
As the junction size increased, the nonlinear inductance decreased, resulting in a potential more harmonic. 
This reduced the accuracy of the plane-wave approximation, thereby necessitating a larger charge space. However, \addt{for flux qubits,} expanding this in terms of the wavefunctions of a harmonic oscillator would require an even larger space, rendering the plane-wave approach more efficient.
Following the convergence trend, to determine circuit parameters with low cost, an initial rough fit \add{is performed using}\erase{employed} $(k, l, m) = (6, 6, 6)$ and \add{a} second fine fit use\erase{d}\add{ing} $(k, l, m) = (9, 6, 7)$, which achieved differences below 2 MHz at all bias points

\begin{figure*}[t]
\centering
\includegraphics[width=180mm]{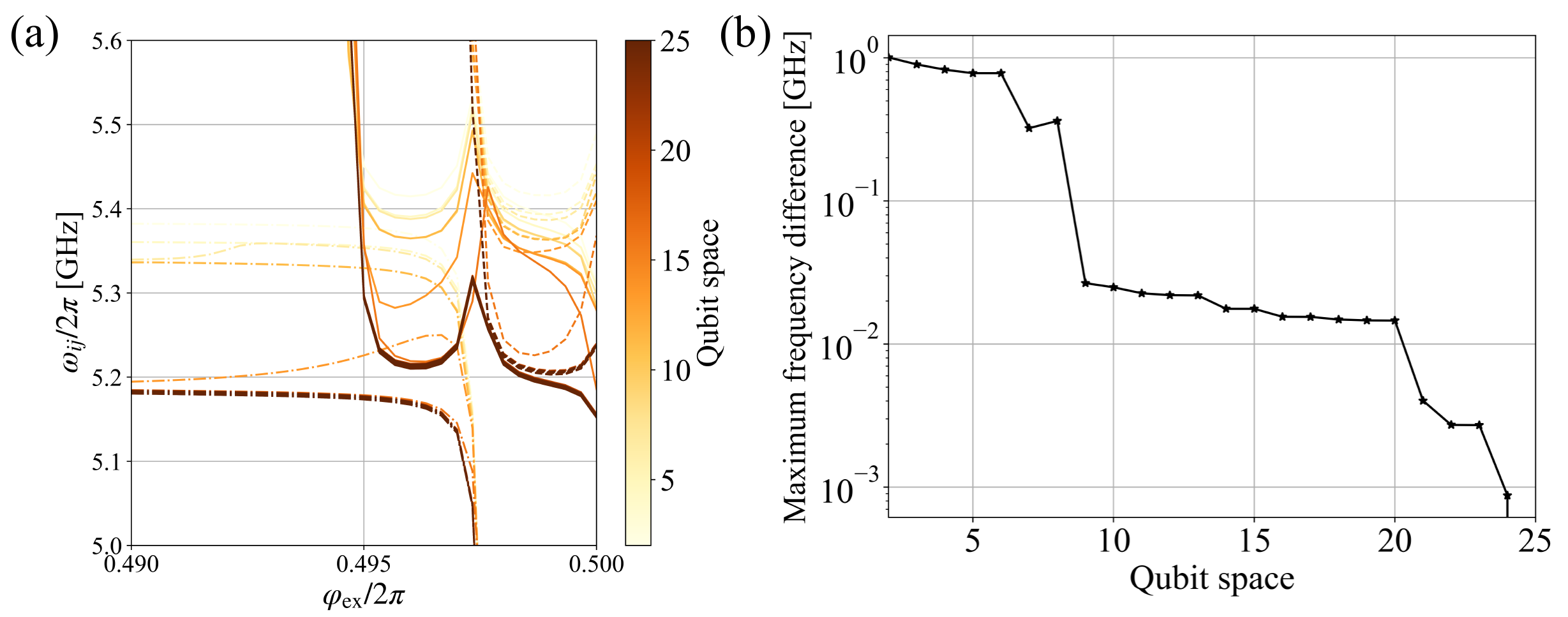}
    \caption{
(a) Spectrum of the qubit space ranging from 2 to 25. When the qubit space \erase{was}\add{is} small, the frequency tended to be higher. Additionally, for certain bias points, there \erase{were}\add{are} regions where an insufficient qubit space failed to adequately represent the system.
(b) Plot of the quantity max\add{\{$\abs{\omega_{10}(Q) - \omega_{10}(25)} + \abs{\omega_{20}(Q) - \omega_{20}(25)} + \abs{\omega_{31}(Q) - \omega_{31}(25)}\}$} within the range $\varphi_\mathrm{ex}/2\pi = 0.49~$–~$0.50$\add{, where Q is the number of qubit space}. 
    }
\label{qspace}
\end{figure*}

Finally, we discuss the qubit space when the coupling term is expanded in the basis using the eigenstates of the qubit Hamiltonian.
Figure~\ref{qspace}(a) shows the spectrum when different numbers of qubit spaces are used with the fitting parameters obtained in Fig.~\ref{Fit}(b). 
In a fashion similar to that of the discussion of the Fock space, the maximum frequency difference between the qubit space sizes of 2–24 and the qubit space size of 25 is plotted in Fig.~\ref{qspace}(b) for $\omega_{10}$, $\omega_{20}$, and $\omega_{31}$).
As the number of expansion terms increased, the spectrum converged (although not monotonically), with the difference reduced to approximately 1 MHz for approximately 23 terms. 
We include\erase{d} up to 25 terms for the fit shown in Fig.~\ref{Fit}(b). The most time-consuming step of the spectrum calculation \erase{was}\add{is} the diagonalization of the qubit. 
Because the inclusion of an additional qubit space after qubit diagonalization \erase{did}\add{does} not significantly affect the computation time, accounting for 25 terms in the actual calculations \erase{was}\add{is} feasible.

For the flux qubit, because the junction inductances at each bias point \erase{were}\add{are} significantly different, the degree of convergence varied according to the bias point.
In the ultrastrong coupling regime, the coupling terms \erase{had}\add{have} values that \erase{could}\add{can} not be ignored even for higher energy levels, resulting in a substantial impact on the eigenvalue calculation of the Hamiltonian. 
As discussed in Ref.~\onlinecite{tomonaga_quasiparticle_2021,yoshihara_hamiltonian_2022}, the resonator frequency tends to be overestimated when the circuit model is approximated using two levels. This occurs because the eigenvalues of the entire Hamiltonian shift upward owing to neglected higher-order coupling (off-diagonal) terms~\cite{masuda_effects_2021}.

\section{Discussion}
In this section, we discuss the validity of the parameters obtained by the fitting.
Using the SEM image shown in Fig.~\ref{Circuit}(b), the area of the $a$-junction \erase{was}\add{is} estimated to be $0.128~\si{\micro m^2}$, with $\alpha = 0.64$ and $\beta = 1.27$. In addition, based on room-temperature resistance measurements, the current density of the Josephson junction was $J_\mathrm{c} \simeq 5~\si{\micro A/\micro m^2}$, resulting in $E_\mathrm{J} \simeq 246$ GHz. The capacitance per unit area of the junction \erase{was}\add{is} approximately $50~\si{fF/\micro m^2}$ under the fabrication conditions and $E_c$ \erase{was}\add{is} approximately $3$ GHz. These values \erase{were}\add{are} reasonably consistent with the fitting parameters.
Regarding the correspondence with the Rabi model parameters, 
$\alpha$ causes $\Delta$ to vary exponentially.
Furthermore, the coupling constant $g$ increased as $\beta$ decreased, and it decreased as $L_\mathrm{r}$ increased. Therefore, once $\beta$ \erase{was}\add{is} reasonably determined from the SEM image, $L_\mathrm{r}$ \erase{was}\add{is} automatically determined, and $C_\mathrm{r}$ \erase{was}\add{is} fixed according to the resonator frequency. By trusting the SEM image and room-temperature resistance measurements, the parameters can be determined almost uniquely~\cite{tomonaga_quasiparticle_2021,miyanaga_ultrastrong_2021}.

\section{Conclusion}
In this study, we described a method for extracting \erase{ridge}\add{valley} structures from nonlinear and complex spectra to efficiently identify peaks. 
Using the example of a complex spectrum in an ultrastrong coupling system, we demonstrated that the peaks are successfully identified. This method is not only efficient for simple spectra but also adaptable to various types of spectra.
\add{Furthermore, in superconducting quantum circuits, low-frequency fluctuations arising from various noise sources are ubiquitous~\cite{burnett_decoherence_2019,schlor_correlating_2019,muller_towards_2019,kristen_giant_2024}. Our method enables the acquisition of qubit parameters with reduced influence from such low-frequency fluctuations by shortening the measurement time.}

We also provided a broadly applicable analysis method for determining the number of expansion terms in numerical calculations for superconducting quantum circuits with multiple Josephson junctions. 

In the peak extraction method, the selection of \erase{ridges}\add{valleys} as transition frequencies involves manual intervention, which introduces subjectivity. To fully automate this process, it is necessary to ensure that the measured spectrum closely matches predictions based on the design parameters or to employ machine learning techniques that can train models on spectrum shapes. 
Automation is essential for processing large-scale circuits and handling multiple lots of data.
In addition, various artificial intelligence-based image processing algorithms are currently being developed, some of which may improve certain aspects of the algorithm proposed in this study (e.g., image sharpening for extracting \erase{ridges}\add{valley}s).
In future work, the validity of the fitting will be further investigated using statistical sampling via the Markov Chain Monte Carlo method ~\cite{metropolis_equation_1953,Hoffman2011TheNS,durmus_convergence_2024}.

\section{Acknowledgement}
We thank K. Inomata and S. Masuda for their thoughtful comments on this study. This work \erase{was}\add{is} supported by the Japan Society for the Promotion of Science (JSPS) KAKENHI (Grant Numbers JP22K21294 and JP23K13048). We would like to thank Editage (www.editage.jp) for English language editing. The programs used in this study are publicly available and can be employed to process various types of experimental data.

\section*{Data availability}
\addt{All data generated or analyzed during this study, as well as the scripts used to produce the figures, are publicly available at https://github.com/AkiyoshiTomonaga/scfitpy.git.}

\subsection*{Competing interests}
\addt{The authors declare no competing interests.}


\bibliography{PT.bib}

\appendix
\renewcommand{\thesection}{Appendix \Alph{section}}
\renewcommand{\theequation}{\Alph{section}\arabic{equation}}
\renewcommand{\thefigure}{A\arabic{figure}}
\setcounter{figure}{0}

\section{Circuit Hamiltonian}\label{sec:A0}
Similar circuits have been introduced in many other studies~\cite{yoshihara_hamiltonian_2022,tomonaga_quasiparticle_2021,forn-diaz_ultrastrong_2017,peropadre_nonequilibrium_2013}. In this study, we also derive the Hamiltonian of the circuit shown in Fig.~\ref{Circuit}(c).
Because the flux qubit has a loop, the junction variables could be reduced using Kirchhoff's law, we eliminate one of the junctions (the $b$-junction).

Starting from branch fluxes across the circuit elements~(the junctions labeled $\beta,\,\alpha,\,a,$ and $b$, the inductance $L_\mathrm{r}$ and capacitance $C_\mathrm{r}$), \eraset{Kirchhoff's voltage laws can be expressed as}\addt{we impose fluxoid quantization only on the loop that carries nonzero persistent current and apply Kirchhoff’s voltage law around the resonator loop~\cite{burkard_circuit_2005,ulrich_dual_2016,vool_introduction_2017}.}
\addt{Thus, we obtain}
\begin{align}
    &\varphi_{\beta}+\varphi_{\alpha}+\varphi_{a}+\varphi_{b} = \varphi_{\mathrm{ex}}\,, \\
    &\varphi_\mathrm{cr}+\varphi_\mathrm{\ell r}+\varphi_{\beta}=0\,, 
    \label{eq:KVL}
\end{align}
where $\varphi_\mathrm{cr}$ and $\varphi_\mathrm{\ell r}$ represent the fluxes between the resonator capacitor and inductor, respectively.
The total Lagrangian of the circuit is defined as the sum of the Lagrangians of the individual circuit component; its formula is
\begin{align}
\mathcal{L}_\mathrm{QR}=\mathcal{L}_\mathrm{Q}+\mathcal{L}_\mathrm{R}+\mathcal{L}_\mathrm{Int}\,,
    \label{eq:Ltot}
\end{align}
where
\begin{align}
    \mathcal{L}_\mathrm{Q} &= \mathcal{K}-\mathcal{U}\,,\\
    \mathcal{K}
    &=\frac{C_\mathrm{J}}{2}[
    \beta\dot{\phi}_{\beta}^2+\dot{\phi}_{a}^2+b(\dot{\phi}_{a }+\dot{\phi}_{\alpha}+\dot{\phi}_{\beta})^2 +\alpha\dot{\phi}_{\alpha}^2] \, 
\end{align}
\begin{align}
        \mathcal{U}=-E_\mathrm{J}
        [
        &\beta\cos(\varphi_{\beta})
        +\cos{(\varphi_{a})}+b\cos{(\varphi_{\mathrm{ex}}-\varphi_a-\varphi_{\alpha}-\varphi_{\beta})} 
        +\alpha\cos{\varphi_\alpha}
        ]+\frac{1}{2L_\mathrm{r}}\phi_{\beta}^2
        \,,
    \label{eq:uj}
\end{align}
\begin{align}
    \mathcal{L}_\mathrm{R}
    =\frac{C_\mathrm{r}}{2}\dot{\phi}_{\mathrm{cr}}^2-\frac{1}{2L_\mathrm{r}}\phi_\mathrm{cr}^2\,,
    \label{Lr}
\end{align}
\begin{align}
      \mathcal{L}_\mathrm{Int}
    =\frac{\phi_\mathrm{cr}\phi_\mathrm{\beta}}{L_\mathrm{r}}\,,
    \label{Lr}  
\end{align}
and $b, \alpha, \beta$ represent the ratios of the areas of those junctions to the size of $a$-junction ($a=1$).
The qubit kinetic energy is written in matrix form as
\begin{align}
    \mathcal{K}= \frac{1}{2} \dot{\boldsymbol{\upphi}}{\!} ^\mathrm{T}\vb{M}\dot{\boldsymbol{\upphi}} \,,
    \label{eq:Kq}
\end{align}
where $\boldsymbol{\upphi} \equiv \mqty(\phi_{\beta} & \phi_{\alpha} & \phi_{a})^\mathrm{T}$ and the capacitance mass matrix is given by
\begin{align}
    \vb{M} &= C_\mathrm{J}\mqty (
        \beta+b & b & b  \\
        b & \alpha+b & b   \\
        b & b & 1+b \\
        ) \,.
        \label{eq:matrix}
\end{align}
Using the canonical conjugate $Q_i= \partial\mathcal{}{L}_\mathrm{tot} / \partial \dot{\phi}_i$ for $\dot{\phi}_i$, where $i\in\{\beta,\alpha,a\}$, we can rewrite Eq.~\eqref{eq:Kq} as
\begin{align}
    \mathcal{K}=\frac{1}{2}\vb{Q}^\mathrm{T}\vb{M}^{-1}\vb{Q}\,,
\end{align}
where $\boldsymbol{Q} \equiv \mqty(Q_{\beta} & Q_{\alpha} & Q_{a})^\mathrm{T}$.
Then we obtain the total Hamiltonian of the circuit as
\begin{align}
   \mathcal{H}_{\mathrm{QR}}&= \vb{Q}^\mathrm{T} \dot{\boldsymbol{\upphi}}-\mathcal{L}_{\mathrm{QR}} \notag\\
    &=\mathcal{H}_\mathrm{Q}+\mathcal{H}_\mathrm{R}+\mathcal{H}_\mathrm{Int} \,,
    \label{eq:TotalHami_A}
\end{align}

In the BCS theory, the superconducting wavefunction is described by $\ket{\Psi_\mathrm{BCS}}=\sum_q\exp{-iq\varphi}\ket{\psi_q}$, where $\ket{\psi_q}$ is the $q$-Cooper pair state vector~\cite{bardeen_theory_1957,tinkham_introduction_2004}.
The relationship between the $q$-Cooper pair state and the phase state $\ket{\psi_\varphi}$ is expressed as
\begin{align}
    \ket{\psi_q}=\frac{1}{2\pi}\int_{-\pi}^{\pi}d\varphi e^{-iq\varphi}\ket{\psi_\varphi}\,.
    \label{wavefunc}
\end{align}
In addition, based on the normalization condition of the BCS state, we obtain
\begin{align}
    \braket{\Psi_\mathrm{BCS}}{\Psi_\mathrm{BCS}}=\sum_{q}\sum_{q'}\exp{i(q'-q)\varphi}\braket{\psi_{q'}}{\psi_q}=1~,
\end{align}
that is, $\braket{\psi_q'}{\psi_q}=\delta_{q,q'}$.
Therefore, we define the basis $\ket{\psi_q}$ as $\ket{q}=(0, 0, ... , 1 (q~\mathrm{th}), ... , 0)^\mathrm{T}$ for the numerical calculations, and the potential terms of the Hamiltonian are expressed in the following matrix form via integrals:
\begin{align}
\hspace{1.4cm}\mel{q'}{\cos{\hat{\varphi}}}{q}&=\mel{q'}{\frac{1}{2}(e^{-i\hat{\varphi}}+e^{i\hat{\varphi}})}{q}\notag\\
&=\qty(\frac{1}{2\pi})^2\int_{-\pi}^\pi\int_{-\pi}^\pi d\varphi^2 e^{-i(q-q')\hat{\varphi}}\frac{1}{2}\qty(e^{-i\hat{\varphi}}+e^{i\hat{\varphi}})\notag\\
&=\frac{1}{2}(\delta_{q',q+1}+\delta_{q',q-1})\,,
\label{cosp}
\end{align}
\begin{align}
\mel{q'}{\varphi}{q}&=\qty(\frac{1}{2\pi})^2\int_{-\pi}^\pi\int_{-\pi}^\pi d\varphi^2 e^{-i(q-q')\hat{\varphi}}\varphi\notag\\
&= \left\{ \begin{array}{l}
\displaystyle \frac{i(-1)^{q-q'}}{q-q'}\,\, (q\neq q')\\\\
\,\,\,\,\,\,\,\,\,\ 0 \,\,\,\,\,\,\,\,\,\,\,\,\,\, (q = q')
\end{array} \right.
\,,
\label{intint}
\end{align}
\begin{align}
\mel{q'}{\varphi^2}{q}&=\qty(\frac{1}{2\pi})^2\int_{-\pi}^\pi\int_{-\pi}^\pi d\varphi^2 e^{-i(q-q')\hat{\varphi}}\varphi \notag\\
&= \left\{ 
\begin{array}{l}
\displaystyle \frac{2(-1)^{q-q'}}{(q-q')^2}\,\,\, (q\neq q')\\\\
\,\,\,\,\,\,\,\,\,\ \pi^\frac{2}{3} \,\,\,\,\,\,\,\,\,\,\,\, (q = q')
\end{array} \right.
\,.
\label{p2int}
\end{align}
The integral in Eq.~\eqref{p2int} is used to calculate the squared term of $\hat{\varphi}_\beta$ in Eq.~\eqref{HQ13}.
Finally, $\mathcal{H}_\mathrm{Q}$ is expressed using a sixth-order tensor:
\begin{align}
\mathcal{H}_\mathrm{Q}^{k,k',l,l',m,m'}=&\langle\psi_{k'}^{(\beta)}|\langle\psi_{l'}^{(\alpha)}|\langle\psi_{m'}^{(a)}|\mathcal{H}_\mathrm{Q}|\psi_k^{(\beta)}\rangle |\psi_l^{(\alpha)}\rangle |\psi_m^{(a)}\rangle \,.
\label{H1tensor}
\end{align}
Similarly, using Eq.~\eqref{intint}, the interacting Hamiltonian $\mathcal{H}_\mathrm{int}$ can be calculated.

\section{Peak trace precision}\label{sec:A1}

\begin{figure*}[t]
\centering
\includegraphics[width=180mm]{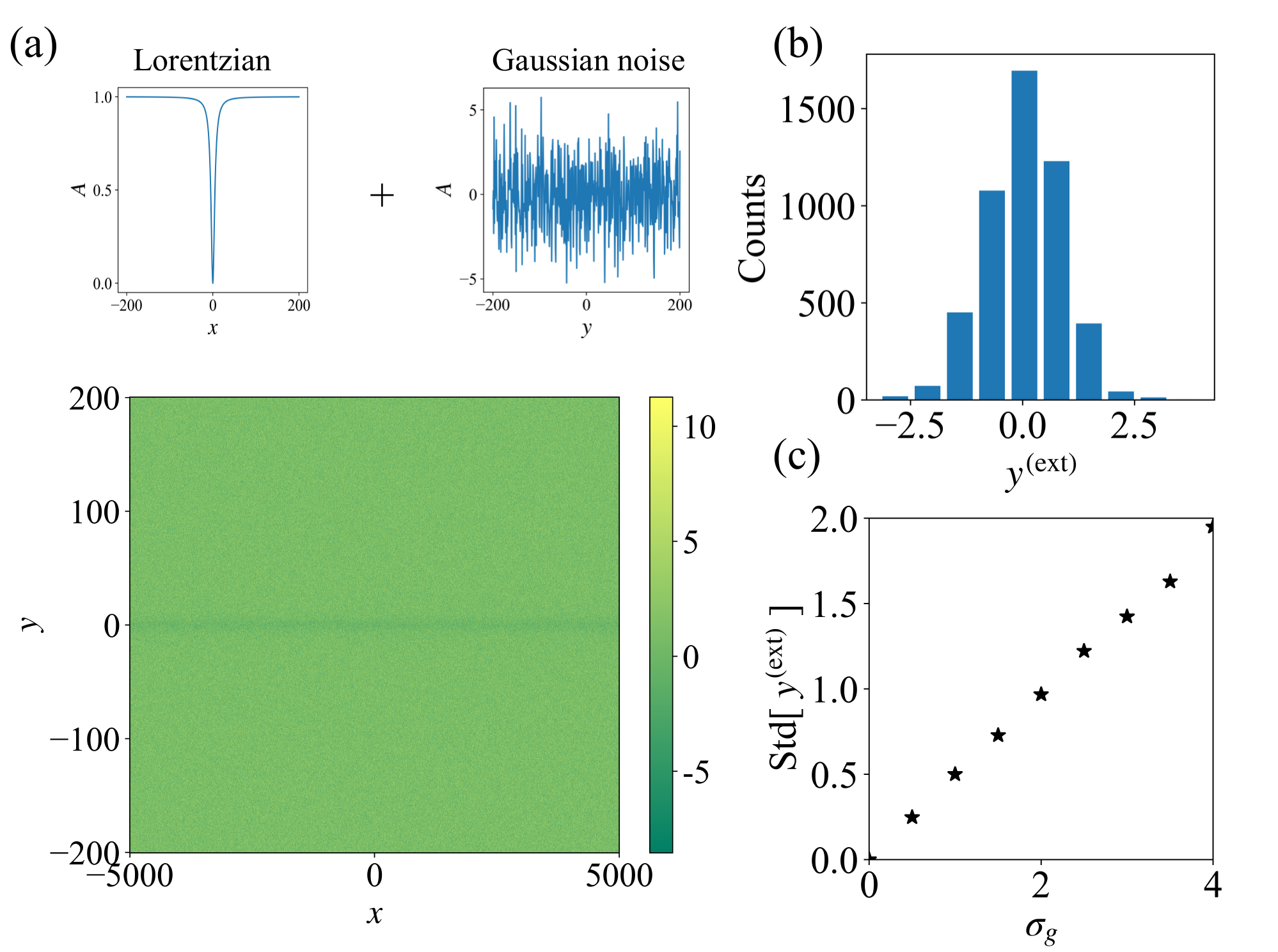}
    \caption{
    (a) A sample spectrum used to verify the stability of peak extraction. 
    A Lorentzian curve is superimposed with Gaussian noise, where the standard deviation $\sigma_g$ of the Gaussian noise is 2.
    (b) The distribution of the extracted peaks, which reflects the Gaussian noise distribution.
    (c) A plot of the standard deviation of the extracted peaks $\sigma_P$ as a function of the standard deviation $\sigma_g$ of the Gaussian noise.
    }
\label{PTnoise}
\end{figure*}

To verify the precision of the peak trace, we prepare\erase{d} a spectrum in which Gaussian noise is added to a Lorentzian signal and perform\erase{ed} peak extraction on $x=10,000$ datasets. 
The amplitude of each attempt ($x$-direction) is $A=L(y)+\mathcal{N}(0,\sigma_g^2)$, where $L(y)=\gamma^2/(4y^2+\gamma^2)$ with $\gamma=10$ and $\mathcal{N}$ is Gaussian noise with strength $\sigma_g^2$.
The extracted peaks follow a Gaussian distribution, as shown in Fig.~\ref{PTnoise}(b), reflecting the noise characteristics. 
When measuring spectra in superconducting circuits using a VNA, Gaussian noise is typically dominant. 
As the strength of Gaussian noise $\sigma_g$ increases, the standard deviation of the extracted peaks also increases linearly, as shown in Fig.~\ref{PTnoise}(c).

\section{Fluxonium}\label{sec:A2}

\begin{figure*}[t]
\centering
\includegraphics[width=180mm]{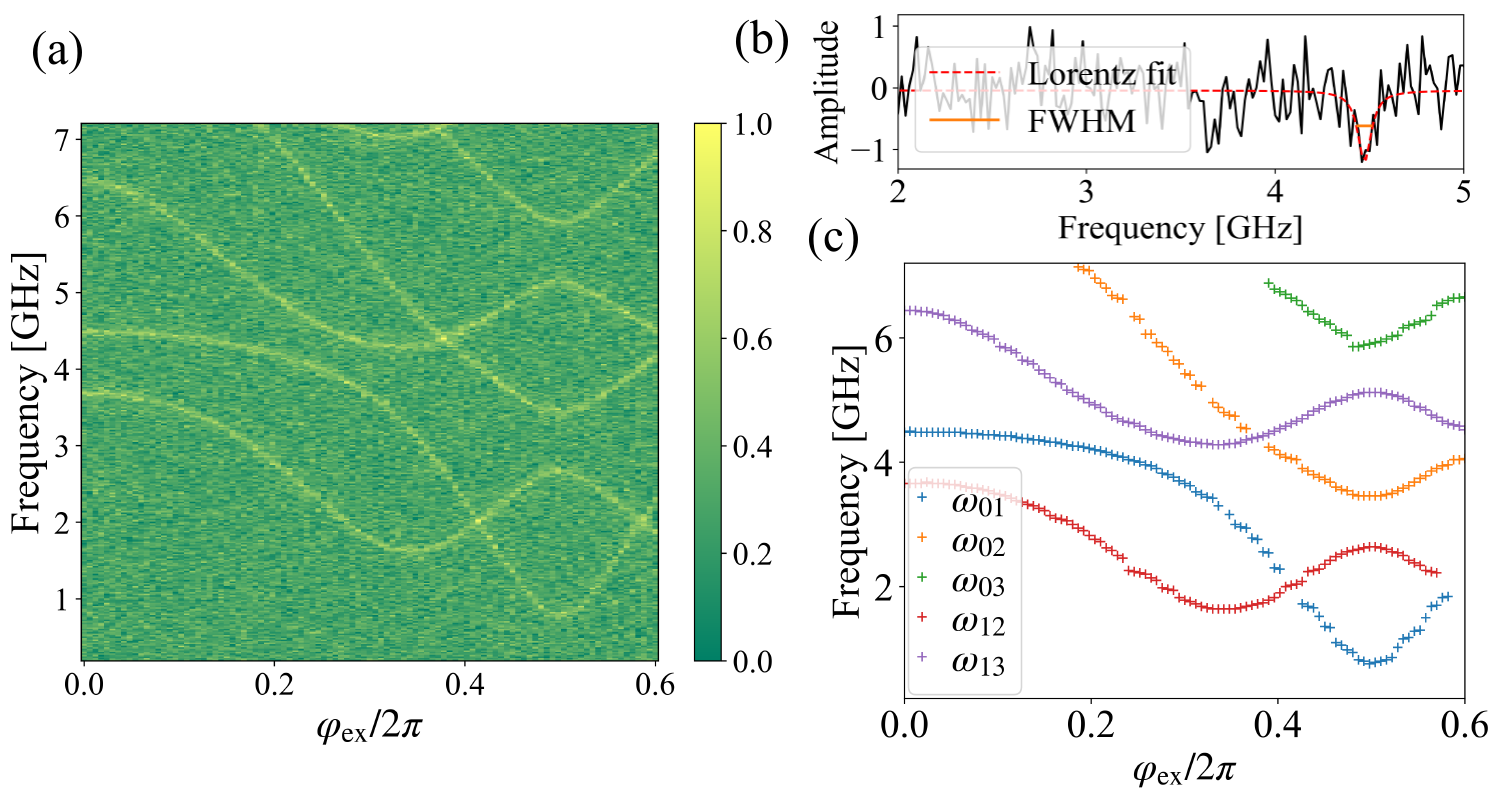}
    \caption{
    (a) Simulated fluxonium spectrum generated with $E_J=3.0$, $E_C=0.84$, and $E_L=0.10\,$GHz. 
    (b) Cross–section of (a) at $\varphi_\mathrm{ex}=0$ and the corresponding Lorentzian fit. 
    (c) Extracted peak traces from (a).
    }
\label{fluxonium}
\end{figure*}

Here we consider the fluxonium example.  Figure~\ref{fluxonium}(a) shows the spectrum generated from the fluxonium Hamiltonian~\cite{nguyen_high-coherence_2019}
\begin{align}
\mathcal{H}_\mathrm{fl}=4E_C\,\hat q^2+\frac{1}{2}E_L\,\hat\varphi^2 - E_J\cos(\hat{\varphi}-\varphi_\mathrm{ex}),
\end{align}
where $E_C$ and $E_L$ denote the Josephson charging energy and the inductive energy of the fluxonium, respectively.  
We use the parameters $E_J=3.0$, $E_C=0.84$, and $E_L=0.10\,$GHz.
For devices such as the transmon or fluxonium whose potential is close to that of a harmonic oscillator, a harmonic–oscillator basis rather than in plane waves as used in the main text should be used for the calculation.
To compute the spectrum of Fig.~\ref{fluxonium}(a), we use the first fifty harmonic–oscillator basis.  

The resonances in the spectrum are generated by placing Lorentzian shape ($\gamma=0.04$) at each transition frequency (the eigenvalue differences of the Hamiltonian), and then adding Gaussian noise with $\sigma_g=0.4$, same as described in Sec.~\ref{sec:A1}.
Although the resulting spectrum in Fig.~\ref{fluxonium}(a) has a different shape than that of a flux qubit, our peak–tracing method still successfully extracts the features, as shown in Fig.~\ref{fluxonium}(c). 
The fluxonium spectrum in Fig.~\ref{fluxonium}(a) has a much wider frequency and bias range than Fig.~\ref{PT}(a), but the peak extraction is well performed and the points traced here are sufficient to fit to obtain the fluxonium circuit parameters.

Finally, we comment on the choice of parameters in the peak tracing algorithm.  In step 1 of our method, the scale factor of the multiscale line filter should match the characteristic linewidth of the peaks to be extracted.  A practical choice is  
\begin{align}
\mathrm{scale~factor}\approx \frac{\mathrm{FWHM}}{2.355},
\end{align}
where a full width half maximum (FWHM) is obtained from a Lorentzian fit and converted to pixel units via the known axis sampling interval as shown in Fig.~\ref{fluxonium}(b). The factor 2.355 comes from the relation between Lorentzian and Gaussian widths.  In a program, the implementation allows passing a list of scale factors, and it will automatically select the one that extracts the most structure.  

In step 2, the contour–length threshold can be set to the pixel length of the smallest feature one wishes to trace, or to a few times the average contour length.  The height threshold should be chosen around 40–80\% of the peak height, depending on the noise level.

\end{document}